\documentclass[sigconf,nonacm]{acmart}
\ifdefined\pdfsuppressptexinfo
\fi

\usepackage{multirow,makecell,colortbl}
\usepackage{stfloats}\usepackage{framed}
\definecolor{hlP}{HTML}{FCE3D3}\definecolor{hlU}{HTML}{D7EFE4}\definecolor{hlKey}{HTML}{EFEFEF}
\definecolor{hlRed}{HTML}{FADBD8}\definecolor{keyred}{HTML}{B3261E}

\newcommand{\bench}{PMRecBench}
\newcommand{\iiagent}{i$^2$Agent}
\newcommand{\DU}{\mathcal{D}_U}
\newcommand{\DP}{\mathcal{D}_P}

\newcommand{\header}[1]{\vspace{1.5mm}\noindent\textbf{#1}.}
\newlength{\dashrulewidth}
\newcommand{\dashedrule}{\noalign{\vskip\aboverulesep\hbox to \dashrulewidth{\xleaders\hbox to 8pt{\vrule height\arrayrulewidth width 4pt\hfil}\hfill}\vskip\belowrulesep}}

\newcommand{\PlaceFigConstruction}{%
\begin{figure*}[t]
  \centering
  \captionsetup{skip=4pt}
  \includegraphics[width=0.95\textwidth,trim=0 0 0 26,clip]{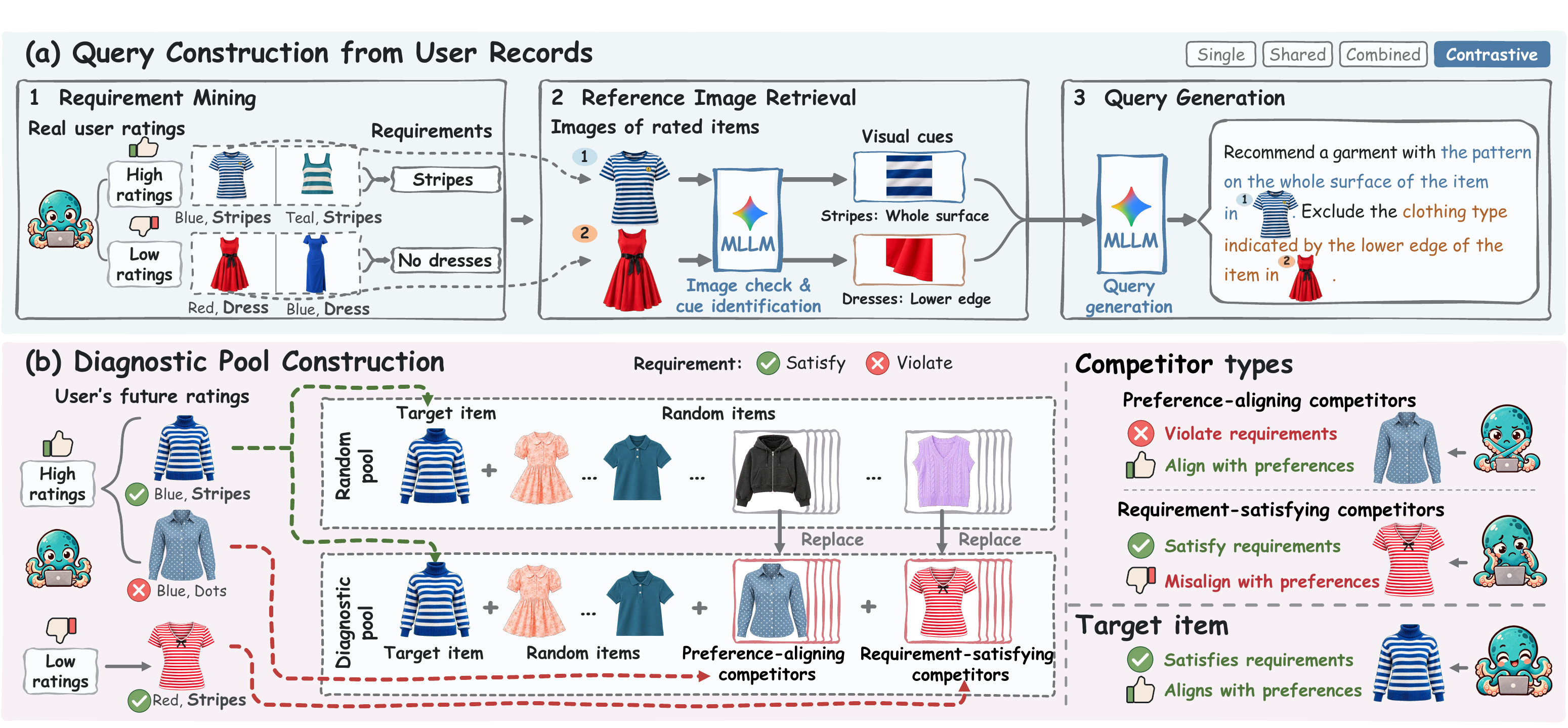}%
  \vspace*{-1mm}
  \caption{Overview of \bench{}: (a) query construction with a Contrastive example and (b) diagnostic pool construction.}
  \label{fig:construction}
  \vspace*{-4mm}
\end{figure*}}

\makeatletter
\g@addto@macro\@mkauthors{%
  \global\setbox\mktitle@bx=\vbox{%
    \unvbox\mktitle@bx
    {\centering\fontsize{11}{13}\selectfont
      \href{https://github.com/zxd-octopus/PMRecBench-Public}{%
        \raisebox{-0.12em}{\includegraphics[height=1em]{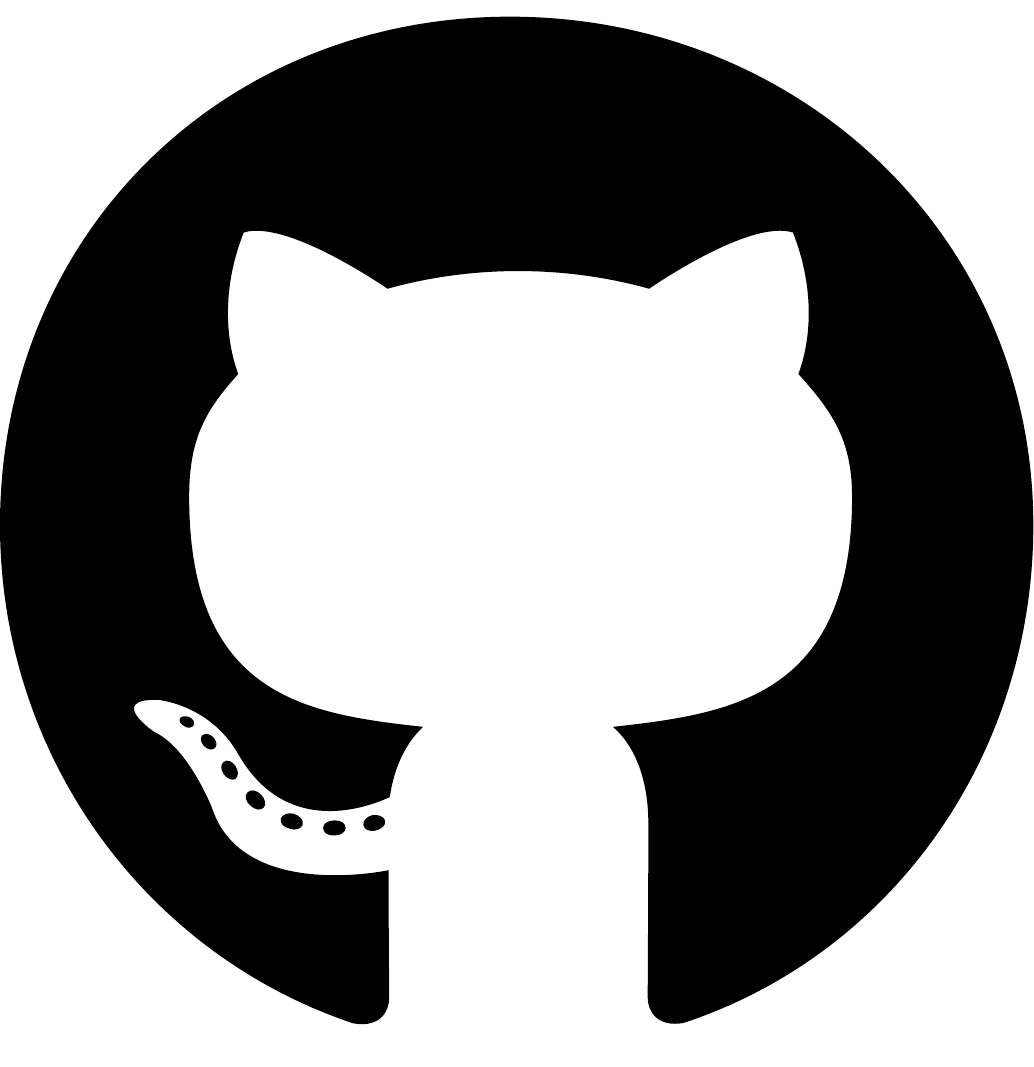}}\,
        \textcolor[HTML]{000080}{Code \& Data}}%
      \hspace{1.5em}%
      \href{https://pmrecbench.github.io/}{%
        \raisebox{-0.12em}{\includegraphics[height=1em]{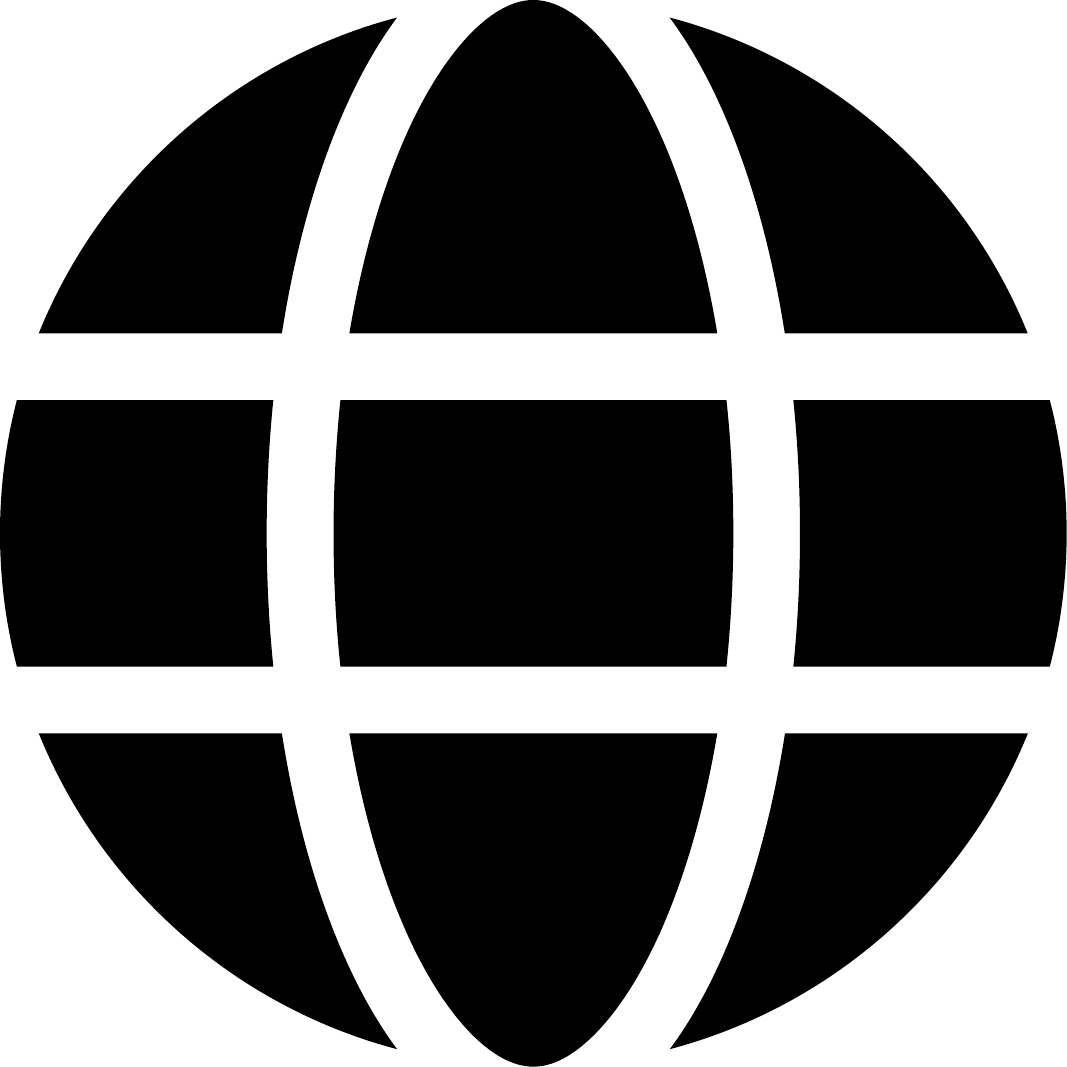}}\,
        \textcolor[HTML]{000080}{Website}}%
      \hspace{1.5em}%
      \href{https://huggingface.co/spaces/tinyoctopus1999/PMRecBench-Leaderboard}{%
        \raisebox{-0.12em}{\includegraphics[height=1em]{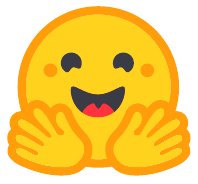}}\,
        \textcolor[HTML]{000080}{Leaderboard}}\par}
    \vspace{1ex}}}
\makeatother

\title{PMRecBench: A Benchmark for Personalization in Recommendation Agents under Multimodal Queries}

\author{Xiaoyu Zhang}
\affiliation{%
  \institution{University College London}
  \city{London}
  \country{United Kingdom}
}
\email{x-zhang.25@ucl.ac.uk}

\author{Yougang Lyu}
\affiliation{%
  \institution{University of Amsterdam}
  \city{Amsterdam}
  \country{The Netherlands}
}
\email{youganglyu@gmail.com}

\author{Bin Wu}
\affiliation{%
  \institution{University College London}
  \city{London}
  \country{United Kingdom}
}
\email{bin.wu.23@ucl.ac.uk}

\author{Sahan Bulathwela}
\affiliation{%
  \institution{University College London}
  \city{London}
  \country{United Kingdom}
}
\email{m.bulathwela@ucl.ac.uk}

\author{Emine Yilmaz}
\affiliation{%
  \institution{University College London}
  \city{London}
  \country{United Kingdom}
}
\email{emine.yilmaz@ucl.ac.uk}

\begin{document}

\begin{abstract}
\begingroup
On Web platforms for shopping and entertainment, recommendation agents under multimodal queries aim to deliver personalized recommendations that satisfy users' visually specified requirements and reflect preferences inferred from user histories.
However, existing personalized recommendation benchmarks use text-only queries, while multimodal ones assess query matching without evaluating personalization based on real user histories.
We introduce the \textbf{P}ersonalized \textbf{M}ultimodal \textbf{Rec}ommendation \textbf{Bench}mark (\bench{}), the first benchmark to evaluate the personalization of recommendation agents under multimodal queries. It pairs 510 quality-controlled image--text queries across clothing and movies with the corresponding users' real histories. \bench{} features two key designs: (1) queries derived from consistent behavioral patterns in real user records, with reference images and targets linked to actual interactions; and (2) challenging diagnostic pools with candidate items rated highly by the user but violating requirements, and items satisfying requirements but never rated highly, testing requirement understanding and preference following, respectively.
Extensive experiments on diagnostic pools reveal a personalization imbalance in recommendation agents: relative to personalization signals alone, adding multimodal queries weakens preference following, even though the additional information about query requirements improves overall ranking performance. Further analysis links this decline in preference following to image interpretation, where agents can prioritize resemblance to reference images over evidence from user histories, and stronger commercial backbones alone fail to resolve this imbalance. This personalization imbalance highlights a central challenge for recommendation agents: improving ranking performance while preserving preference following when responding to multimodal user queries.
\par\endgroup
\end{abstract}



\maketitle

\section{Introduction}
\label{sec:intro}

\begingroup

Large language models (LLMs) are enabling recommendation agents~\cite{huang2025interecagent,xu2025iagent,tang2025recbot} to interpret natural-language requests~\cite{zhang2024empathetic} and provide personalized recommendations~\cite{zhang2023variational}. These agents draw on users' past item interactions and feedback to infer preferences that may not be stated in the current request~\cite{cai2025personalwab}. Advances in multimodal LLMs (MLLMs) support extending these interactions to image--text queries~\cite{bai2025qwen3vl,gemmateam2026gemma4}. For example, a user can ask for clothing with the pattern shown in a reference image. Such image--text queries, already supported by Web shopping interfaces such as Amazon's Alexa for Shopping~\cite{amazon2026alexa}, have become a new way to interact with recommendation services and are also explored in recommendation research~\cite{yoon2024imagery,hao2026mmshopbench}. Personalization in this setting therefore requires agents to meet the current image--text requirements while actively accounting for preferences inferred from the user's interaction history~\cite{siro2025humanllm}.

\PlaceFigConstruction
However, existing benchmarks offer limited evidence on whether agents follow implicit preferences reflected in users’ interaction histories when responding to multimodal queries. Representative personalized recommendation benchmarks evaluate preference-aligned item selection under textual queries~\cite{yang2026apeb,huang2026recbench}. Multimodal queries make this selection harder: agents must first interpret the requirements conveyed by reference images, and it remains unclear how agents combine preference modeling with this interpretation. Conversely, multimodal benchmarks assess image--text query matching without testing whether recommendations reflect the preferences in real user histories~\cite{yoon2024imagery,hao2026mmshopbench}. Yet multiple items may satisfy the same query while differing in their suitability for a particular user. Assessing personalization in this setting calls for queries whose text, reference images, and combination are grounded in real user behavior, so that the query introduces no requirement that conflicts with this user’s preferences. Even with such queries, randomly selected candidates, commonly used in recommendation evaluation~\cite{krichene2020sampled}, may reward coarse query--item matching without adequately testing preference following~\cite{yang2026apeb}. Overall ranking metrics alone also cannot distinguish requirement understanding from preference following. Evaluation therefore needs challenging candidates and separate proxy measures for these two abilities.

We introduce the \textbf{P}ersonalized \textbf{M}ultimodal \textbf{Rec}ommendation \textbf{Bench}mark (\bench{}) to evaluate recommendation agents' personalization under image--text queries. Two complementary designs address these evaluation needs: (1) We derive query requirements from consistent behavioral patterns in real user records, with reference images and targets linked to actual interactions. (2) We construct challenging diagnostic candidate pools that pair a query-satisfying target rated highly by the user with two competitor groups: highly rated items that violate the query, and query-satisfying items the user disliked or never interacted with. The latter serve as proxies for preference misalignment. Two separate diagnostic metrics use the target's rankings relative to these groups as proxies for requirement understanding and preference following, respectively. \bench{} pairs 510 image--text queries with the corresponding users' real histories from Amazon and MovieLens, covering physical products (clothing) and digital content (movies), respectively. Ranking candidate items therefore requires agents to interpret reference images for requirement understanding and infer user preferences from user histories.

Our extensive experiments reveal a personalization imbalance in recommendation agents: \textcolor{black}{adding multimodal queries to personalization signals weakens preference following, even though the additional information about query requirements improves overall ranking performance.} This imbalance is masked by strong results on random candidate pools, where coarse query matching can often suffice. Specifically, ablations on diagnostic candidate pools show that agents follow preferences well with personalization signals alone, but adding multimodal queries lowers the mean preference-following score across four agents by approximately 4–8 percentage points, depending on the domain and backbone. Comparisons of query representations link this decline to image interpretation, while sampled failure traces show that agents can prioritize query-based cues, such as resemblance to reference images, over personalization signals. Upgrading the backbone alone does not resolve the imbalance: with the agent fixed to ReAct, the tested commercial models follow preferences less well than Gemma-4-31B despite comparable or stronger requirement understanding. These findings indicate that agents' demonstrated ability to use personalization signals is not consistently reflected in their final recommendations, a failure mode relevant to Web platforms serving image--text input.

Our contributions are summarized as follows:
\vspace*{-2mm}
\begin{itemize}
    \item We introduce \textbf{\bench{}}, a benchmark for evaluating recommendation agents' personalization under multimodal queries, pairing real user histories with image--text requests grounded in consistent behavioral patterns and reference images of items the users have rated.
    \item We develop \textbf{a diagnostic evaluation protocol} that uses two groups of challenging competitors to test requirement understanding and preference following jointly, with separate proxy metrics for each capability.
    \item We reveal \textbf{a personalization imbalance}: \textcolor{black}{adding multimodal queries to personalization signals weakens preference following, even though the additional information about query requirements improves overall ranking performance.} Our analysis links this decline to image interpretation.
\end{itemize}
\vspace*{-2mm}
\endgroup

\section{\bench{}}
\label{sec:benchmark}
\begingroup
To evaluate recommendation agents' personalization, we construct multimodal queries and diagnostic candidate pools for \bench{} (Figure~\ref{fig:construction}). First, we derive requirements from consistent patterns in users' ratings and express them through text and reference images of items they have rated. Each query has a highly rated target that satisfies its requirements. Second, we place this target in a candidate pool with highly rated items that violate the query and query-satisfying items that the user never rated positively. The target's relative rankings against these two groups provide separate proxies for requirement understanding and preference following.

\subsection{Problem Formulation}
\label{sec:task}
Each test case asks a method to rank a candidate pool $C=\{c_i\}_{i=1}^{N}$ for user $u$ using a multimodal query $Q=(x,\mathcal{I})$ and historical personalization signals $S=(H_u,s_u)$~\cite{hou2024llmrank}. Here, $x$ is the query text, $\mathcal{I}$ is an ordered list of reference images, $H_u$ contains the history items with their ratings, and $s_u(c)$ is the collaborative-filtering score of each candidate item $c$. Each query $Q$ is associated with $m$ requirements $\mathcal{K}_Q=\{\kappa_j\}_{j=1}^{m}$. Each requirement $\kappa_j=(a_j,\sigma_j)$ specifies an attribute $a_j$ to match ($\sigma_j=+$) or avoid ($\sigma_j=-$), and $A(i)$ denotes the attribute set of item $i$.
Therefore, a test case is stored as $(Q,S,C,y^*)$, where the target $y^*\in C$ received a high rating from the user and satisfies every requirement: $a_j\in A(y^*)$ for positive requirements and $a_j\notin A(y^*)$ for negative ones. The method must output a ranking $\pi$ of all $N$ candidates by both whether they satisfy the requirements and whether they align with the user's preferences.

\vspace*{-1mm}
\subsection{Query Construction from User Records}
\label{sec:construction}
To ground each query in real user behavior, we build each test case from real user records. \bench{} uses datasets of two Web platforms, Amazon clothing reviews~\cite{ni2019amazon} and MovieLens-1M~\cite{harper2015movielens}, which provide user ratings and allow item images to be obtained, as Figure~\ref{fig:construction}(a) illustrates with an example. Construction follows four steps. \emph{Requirement mining} identifies attributes~\cite{lyu2022legaljudgment} consistently associated with a user's likes or dislikes. \emph{Reference image retrieval} finds images of items the user rated that visibly show those attributes. \emph{Query generation} turns the attributes and images into natural recommendation requests. \emph{Query validation} checks that the queries express the intended requirements, have to be resolved using the images, and do not reveal the target.

We split each user's chronologically ordered rating records into history (the earlier 70\%) and future ratings (the remaining 30\%). Queries are constructed from history items, while targets are highly rated items from the future ratings.

\subsubsection{Requirement Mining}
\label{sec:condition-mining}
Requirement mining links queries to recurring patterns in a user's ratings. We use item attributes as requirements. For clothing, item attributes $A(i)$ come from MAVE~\cite{yang2022mave}. For movies, we use the Tag Genome~\cite{vig2012taggenome}, TMDB\footnote{\url{https://www.themoviedb.org}}, and Wikidata~\cite{vrandecic2014wikidata}. We use three classes of image-visible attributes: concrete visual details, abstract visual styles, and recognizable entities. Treating items rated at least 4 as liked and those rated at most 2 as disliked, we require users to have sufficient items in both groups, with eligibility thresholds given in Appendix~\ref{app:data-sources}.

Following RecBench+~\cite{huang2026recbench}, we identify attributes shared by multiple distinct rated items as evidence of recurring behavior. An attribute $a_j$ becomes a requirement $\kappa_j=(a_j,\sigma_j)$: positive when shared items tend to be liked, and negative when they tend to be disliked. Attributes used for negative requirements must be more common among disliked items than among liked items. Pairing each requirement set with a target $y^*$ produces 800 pairs of requirement sets and target items. History items carrying each mined attribute serve as a source for reference image retrieval.

\subsubsection{Reference Image Retrieval}
\label{sec:image-retrieval}
Each reference image must meet two conditions: its source item has a user rating consistent with the requirement's polarity and the required attribute is visible in the image. For each requirement $\kappa_j$, we retrieve product images for clothing and web-searched stills for movies of qualifying items from the user's history.

We have Gemini~3.1~Pro check attribute visibility from the image. An image is accepted only if the model identifies the attribute's visual cues and their corresponding image regions and reports a confidence score of at least 0.75. For abstract visual styles, the model must additionally identify supporting cues in at least three distinct regions. These cues and regions serve as location anchors for query generation. Additionally, movie stills must contain no readable text, reducing textual clues about the attribute. If any requirement lacks an accepted image we discard that requirement-set--target pair. This step leaves 673 pairs with accepted reference images~$\mathcal{I}$.

\subsubsection{Query Generation}
\label{sec:query-generation}
\begin{table}[t]
\caption{Statistics of \bench{}.}
\label{tab:stats}
\vspace*{-2mm}
\centering\footnotesize\setlength{\tabcolsep}{2.5pt}
\begin{tabular*}{\linewidth}{@{\extracolsep{\fill}}lrr@{\hspace{8pt}}lrr@{}}
\toprule
\textit{Data size} & Clothing & Movies & \textit{Queries per form} & Clothing & Movies \\
\cmidrule(r){1-3}\cmidrule(l){4-6}
Queries & 258 & 252 & Single & 67 & 65 \\
Reference images & 438 & 439 & Shared & 61 & 62 \\
Users & 198 & 179 & Combined & 62 & 61 \\
Interacted items & 13,511 & 3,123 & Contrastive & 68 & 64 \\
\bottomrule
\end{tabular*}
\vspace*{-5mm}
\end{table}

We use the requirements and accepted reference images to construct four forms of multimodal queries.

\header{Query forms} A \emph{Single} query uses one image showing a positive requirement. A \emph{Shared} query requires accepted images of two different previously rated items showing the same positive requirement; its text describes the requirement vaguely, so the agent needs to infer it from visual cues shared by both images. A \emph{Combined} query uses images showing two distinct positive requirements. A \emph{Contrastive} query pairs one image showing a positive requirement with another showing a negative requirement.

\header{Drafting and paraphrasing} We use Gemini~3.1~Pro to generate the query text $x$ through \emph{drafting} and \emph{paraphrasing}, preserving the need to interpret the reference images while improving naturalness and diversity. During offline drafting, the model receives the reference images $\mathcal{I}$, their requirements and location anchors, and the target item with its images. It writes a recommendation query specifying which attribute type to match or avoid in each reference image. The text may locate an attribute only through the supplied anchors; it must leave attribute values to be inferred from the reference images and must not describe the target item. During paraphrasing, the same model rewrites each draft five times while preserving its requirements. Two judges, Claude Opus~4.8 and GPT-5.5, score each paraphrase~\cite{sauter2026evoideator} from 1 to 5 on recommendation intent and naturalness, and we use the highest-scoring paraphrase.

\subsubsection{Query Validation}
\label{sec:quality}
We validate the 673 generated queries with Claude Opus~4.8 and GPT-5.5 as judges.

\header{Validation checks} Three checks assess whether queries preserve the intended requirements and require the reference images for interpretation without permitting text-only shortcuts. First, we check that the query neither describes the target item nor introduces requirements beyond those specified. Second, the \emph{requirement-clarity check} tests whether each judge can recover every requirement from the query text and reference images through multiple-choice questions. Third, the \emph{text-leakage check} tests whether the text alone reveals requirements without visual evidence.

\header{Agreement and human review} Across the requirement-clarity and text-leakage checks, inter-judge agreement is 0.82 in Gwet's AC1~\cite{gwet2008ac1}. We remove queries that both judges identify as failing any of the three validation checks. The authors review disputed queries~\cite{guo2026rubricreviewer} using the same checks and remove those that fail. On 50 randomly sampled queries, human--judge agreement is 0.78 in Gwet's AC1. After validation, 510 queries remain: 258 for clothing and 252 for movies (Table~\ref{tab:stats}).

\vspace*{-2mm}
\subsection{Evaluation Protocol and Metrics}
\label{sec:evaluation}
Following MSRBench~\cite{zhou2025msrbench}, we construct a random candidate pool of 30 candidates: the target item $y^*$ and 29 items that the user has never interacted with. For clothing, the sampled items share the target's fine-grained category to avoid trivial category matching. Random sampling may nevertheless reward coarse query--item matching and provide limited evidence about preference following. We therefore construct a diagnostic pool with two groups of competitors for more targeted comparisons. Following prior work~\cite{hou2024llmrank}, we use normalized discounted cumulative gain (NDCG)~\cite{jarvelin2002ndcg} to measure overall ranking performance. On the diagnostic pool, we additionally propose two new metrics $U$ and $P$ as relative-ranking proxies for requirement understanding and preference following, respectively (Section~\ref{sec:metrics}).

\subsubsection{Diagnostic Pool Construction}
\label{sec:candidates}
The diagnostic pool pairs the target with two groups of competitors for evaluating requirement understanding and preference following (Figure~\ref{fig:construction}(b)). It replaces ten random candidates with five competitors from each group, yielding 30 items: the target, 19 random candidates, and 10 competitors. For clothing, all of these competitors also come from the same category as the random candidates.

\header{Preference-aligning competitors} Items in $\DU$ were rated highly by the user in the later 30\% of records, like the target, but each lacks at least one required positive attribute or, for a Contrastive query, contains the negative attribute to avoid. Hence, only by checking the requirements can an agent rank the target above these competitors.

\header{Requirement-satisfying competitors} Items in $\DP$ satisfy every query requirement, as does the target. They are items that the user disliked in the later 30\% of records; if fewer than five such items exist, never-interacted items fill the remaining slots as negatives, following prior evaluation protocols~\cite{he2017ncf,kang2018sasrec}. To limit the chance of matching user preferences, each filled item's overlap with combined attributes from history items rated 3 or higher cannot exceed the target's overlap by more than one. Ranking the target above these requirement-satisfying competitors is therefore a proxy for preference following.

\header{Attribute verification} We verify each item's attributes against the requirements before selecting competitors. For $\DU$, an item counts as lacking a positive attribute only if neither a hand-built synonym table nor a DeBERTa-v3-based natural language inference (NLI) model~\cite{he2023debertav3} finds a semantically similar term in its attribute labels or title. The presence of a negative attribute is checked by exact matching. For $\DP$, all requirements are checked by exact matching, and the attribute of negative polarity must also be safely missing by the same test as for $\DU$. In both pools, the target is the only candidate that satisfies all query requirements and has an observed high rating from the user.

\subsubsection{Evaluation Metrics}
\label{sec:metrics}
\begin{table*}[t]
\caption{Main results on the random pool of the clothing and movie domains. We report NDCG@1, @3, and @5 for all methods. Best results are in \textbf{bold}; second best are \underline{underlined}. Conventional recommenders do not depend on the backbone, so their results are identical across backbones.}
\label{tab:main}
\vspace*{-2mm}
\centering\footnotesize
\renewcommand{\arraystretch}{0.8}\setlength{\aboverulesep}{0.3ex}\setlength{\belowrulesep}{0.4ex}
\setlength{\tabcolsep}{0pt}
\setlength{\dashrulewidth}{0.86\textwidth}
\begin{tabular*}{\dashrulewidth}{@{\extracolsep{\fill}}l ccc ccc ccc ccc}
\toprule
\multirow{3}[4]{*}{\textbf{Method}} & \multicolumn{6}{c}{\textbf{Clothing}} & \multicolumn{6}{c}{\textbf{Movies}} \\
\cmidrule(lr){2-7}\cmidrule(lr){8-13}
& \multicolumn{3}{c}{Gemma-4-31B} & \multicolumn{3}{c}{Qwen3.6-35B} & \multicolumn{3}{c}{Gemma-4-31B} & \multicolumn{3}{c}{Qwen3.6-35B} \\
\cmidrule(lr){2-4}\cmidrule(lr){5-7}\cmidrule(lr){8-10}\cmidrule(lr){11-13}
& @1 & @3 & @5 & @1 & @3 & @5 & @1 & @3 & @5 & @1 & @3 & @5 \\
\midrule
\multicolumn{13}{l}{\textit{\textbf{Conventional recommenders}}} \\
SASRec       & 0.155 & 0.224 & 0.265 & 0.155 & 0.224 & 0.265 & 0.413 & 0.558 & 0.596 & 0.413 & 0.558 & 0.596 \\
VBPR         & 0.190 & 0.322 & 0.384 & 0.190 & 0.322 & 0.384 & 0.444 & 0.628 & 0.658 & 0.444 & 0.628 & 0.658 \\
Popularity   & 0.202 & 0.306 & 0.338 & 0.202 & 0.306 & 0.338 & 0.230 & 0.344 & 0.378 & 0.230 & 0.344 & 0.378 \\
Random       & 0.033 & 0.071 & 0.098 & 0.033 & 0.071 & 0.098 & 0.033 & 0.071 & 0.098 & 0.033 & 0.071 & 0.098 \\
\dashedrule
\multicolumn{13}{l}{\textit{\textbf{Direct rankers}}} \\
Direct MLLM  & 0.411 & 0.527 & 0.554 & 0.310 & 0.404 & 0.439 & 0.456 & 0.608 & 0.630 & 0.425 & 0.524 & 0.542 \\
Text-only    & 0.174 & 0.257 & 0.305 & 0.178 & 0.241 & 0.268 & 0.254 & 0.354 & 0.383 & 0.190 & 0.275 & 0.303 \\
MLLM-CoT     & 0.457 & 0.585 & 0.612 & 0.411 & 0.548 & 0.586 & 0.571 & 0.726 & 0.742 & 0.540 & 0.669 & 0.699 \\
\dashedrule
\multicolumn{13}{l}{\textit{\textbf{Domain-general agents}}} \\
ReAct        & \textbf{0.578} & \textbf{0.681} & \textbf{0.712} & 0.492 & 0.613 & 0.660 & 0.659 & 0.781 & 0.799 & 0.611 & 0.711 & 0.745 \\
PyVision     & 0.508 & 0.633 & 0.672 & 0.488 & 0.599 & 0.640 & \textbf{0.683} & \textbf{0.798} & \textbf{0.819} & 0.651 & 0.761 & 0.775 \\
\dashedrule
\multicolumn{13}{l}{\textit{\textbf{Recommendation-specific agents}}} \\
RecMind      & 0.543 & 0.665 & 0.701 & \textbf{0.519} & \textbf{0.641} & 0.660 & \underline{0.667} & 0.775 & 0.793 & 0.595 & 0.731 & 0.752 \\
InteRecAgent & \underline{0.550} & 0.653 & 0.705 & \underline{0.512} & \underline{0.636} & \textbf{0.672} & 0.647 & 0.780 & 0.800 & \underline{0.675} & \underline{0.780} & \underline{0.801} \\
MACRec       & 0.516 & 0.634 & 0.670 & 0.508 & 0.630 & \underline{0.666} & 0.635 & 0.759 & 0.771 & \textbf{0.687} & \textbf{0.813} & \textbf{0.828} \\
\iiagent{}   & 0.547 & \underline{0.680} & \underline{0.707} & 0.496 & 0.617 & 0.648 & 0.635 & \underline{0.786} & \underline{0.801} & 0.667 & 0.779 & 0.794 \\
\bottomrule
\end{tabular*}
\vspace*{-3.5mm}
\end{table*}

NDCG summarizes overall recommendation performance but does not distinguish which group of competitors outranks the target. We therefore introduce the \textbf{\textit{requirement-understanding score}} ($U$) and \textbf{\textit{preference-following score}} ($P$) on the diagnostic pool. Let $\mathcal{T}$ be the set of test cases. In test case $t$, $y_t^*$ is the target and $\mathcal{D}_{g,t}$ is the set of its five competitors in group $g$, where $g=U$ denotes the requirement-violating competitors and $g=P$ the preference-misaligning competitors. Let $r_t(i)$ denote item $i$'s position in the output ranking $\pi_t$, with smaller values indicating higher ranks. Pooling all target--competitor pairs gives
\begin{equation}
\begin{aligned}
M_g &= \frac{\sum_{t\in\mathcal{T}}\sum_{d\in\mathcal{D}_{g,t}}\mathbf{1}[r_t(y_t^*)<r_t(d)]}{\sum_{t\in\mathcal{T}}|\mathcal{D}_{g,t}|},\\
U &= M_U,\qquad P=M_P.
\end{aligned}
\label{eq:diagnostic-metrics}
\end{equation}
{Here, $\mathbf{1}[\cdot]$ is 1 when its condition holds and 0 otherwise, and the denominator counts all comparison pairs across test cases. $U$ compares the target with high-rated items that violate a requirement; $P$ compares it with items that satisfy all requirements but the user never rated positively. Higher scores mean the target more often outranks corresponding competitors. These scores are ranking-based proxies for requirement understanding and preference following.\par}
\par\endgroup

\vspace*{-2mm}
\section{Experiments}
\label{sec:setup}

\subsection{Research Questions}
\label{sec:rq}
\begingroup

Our experiments answer the following research questions:
\begin{itemize}
\item \textbf{RQ1:} How well do existing methods handle personalized recommendation under multimodal queries?
\item \textbf{RQ2:} Do agents preserve preference following when they process multimodal queries?
\item \textbf{RQ3:} How does the representation of the multimodal query affect preference following?
\item \textbf{RQ4:} Do stronger backbone models improve preference following?
\item \textbf{RQ5:} Which decision steps lead agents to fail?
\end{itemize}
\par\endgroup

\vspace*{-2mm}
\subsection{Baselines}
\label{sec:baselines}
The primary goal of \bench{} is to test whether existing systems can satisfy a multimodal query while following the user's preferences. To this end, we evaluate representative recommendation methods and backbone models of different types and scales.

{%
\header{Methods} The methods fall into four groups, ordered from the simplest to the most specialized. (1)~\textbf{Conventional recommenders} do not read the query and rank the candidates only from the users' history: SASRec~\cite{kang2018sasrec} models the order of the history items with self-attention, VBPR~\cite{he2016vbpr} adds item-image features to matrix factorization, Popularity ranks the candidates by the number of interactions with each candidate, and Random gives the expected score of a random ranking. (2)~\textbf{Direct rankers}, following LLMRank~\cite{hou2024llmrank}, let an MLLM rank all candidates from the user's history, the query, and the candidates in one prompt without calling tools: Direct MLLM receives the query text and the reference images of the query, Text-only removes the reference images and keeps only the query text, and MLLM-CoT receives the same input as Direct MLLM plus a zero-shot chain-of-thought instruction~\cite{kojima2022cot} to reason before ranking~\cite{lyu2023multidefendant}. (3)~\textbf{Domain-general agents} are not designed for recommendation~\cite{wu2026agenttoolsearch}: ReAct~\cite{yao2023react} interleaves reasoning~\cite{shi2025deepresearch} with tool calls~\cite{kanoulas2025agentcentric}, and PyVision~\cite{zhao2025pyvision}, designed specifically for multimodal reasoning, lets an MLLM write and run Python tools. (4)~\textbf{Recommendation-specific agents} cover common paradigms of reasoning and tool use for recommendation: RecMind~\cite{wang2024recmind} plans step by step with a Self-Inspiring algorithm; InteRecAgent~\cite{huang2025interecagent} follows plan-first execution over recommender tools; MACRec~\cite{wang2024macrec} coordinates specialized agents under a manager; and \iiagent{}~\cite{xu2025iagent} runs a fixed workflow that maintains a dynamic memory of each user's feedback. Following PersonalWAB~\cite{cai2025personalwab}, all agents and direct rankers use Gemma-4-31B and Qwen3.6-35B as backbone models in Sections~\ref{sec:overall}--\ref{sec:representation}, so that performance differences reflect the methods.\par}

\header{Backbone Models} We evaluate eight MLLMs, including (1)~\emph{small open-weight models} with fewer than 10B parameters, Qwen3-VL-8B~\cite{bai2025qwen3vl}, Ministral~3~8B~\cite{liu2026ministral3}, and Gemma-4-E4B~\cite{gemmateam2026gemma4}; (2)~\emph{medium open-weight models} of about 30B parameters, Gemma-4-31B~\cite{gemmateam2026gemma4} and Qwen3.6-35B~\cite{qwen2026qwen36}; and (3)~\emph{commercial models}, GPT-6 Luna~\cite{openai2026gpt6luna}, GLM-5.3-Flash~\cite{zai2026glm53flash}, and Claude Haiku~5.5~\cite{anthropic2026haiku55}. The three groups cover open-weight and commercial models of different scales, so that Section~\ref{sec:backbone} can measure, with ReAct as the method, how backbones with different capabilities affect satisfying the query and following the user's preferences.

\vspace*{-1mm}
\subsection{Implementation Details}
\label{sec:implementation}
We add the task instruction of Section~\ref{sec:task} to the original prompt of each agent and direct ranker (Appendix~\ref{app:tools}). Gemma-4-31B and Qwen3.6-35B decode with temperature 0.
As in AgentRecBench~\cite{shang2025agentrecbench}, the tool-calling agents additionally share one tool set; ours has seven tools that return rated history items, item profile, item reviews (clothing only), collaborative-filtering scores, item images, image captions, and CLIP~\cite{radford2021clip} similarities to the reference images.

\vspace*{-1mm}
\section{Experimental Results and Analysis}
\label{sec:results}

To answer our research questions, we compare all methods on the random pool, contrast inputs and query representations on the diagnostic pool, compare backbones with ReAct fixed, and analyze the decision steps behind failures. Due to space constraints, the Qwen3.6-35B results of Sections~\ref{sec:balance} and~\ref{sec:representation} appear in Appendix~\ref{app:results} and support the same conclusions.

\vspace*{-1mm}
\subsection{Main Results (RQ1)}
\label{sec:overall}

\begingroup

{Table~\ref{tab:main} compares agents and direct rankers, running on Gemma-4-31B and Qwen3.6-35B, with conventional recommenders on the random pool. Both reference images and the user's history carry information about the target, and agents, which combine the two through multi-step tool use, perform best. We have four observations:\par}

{\header{Agents outperform direct rankers and conventional recommenders} In every setting, even the weakest agent scores higher than the strongest direct ranker and the strongest conventional recommender, and on average the six agents exceed the three direct rankers in NDCG@1 by 0.20 on clothing and 0.25 on movies, a significant gap in every setting (95\% bootstrap interval). None of the four recommendation-specific agents differs significantly from ReAct or PyVision on either backbone in either domain. Since agents must match both the multimodal query requirements and the user preference to identify each target in \bench{}, gathering and combining both kinds of evidence matters more than recommendation-specific designs.\par}

\header{Removing the reference images sharply lowers ranking accuracy} As shown in Table~\ref{tab:main}, Text-only, which keeps only the query text, reaches an NDCG@1 of 0.17 to 0.25, significantly below the 0.31 to 0.46 of Direct MLLM, which also receives the reference images. Because the query text in \bench{} names only the attribute type and leaves its value to the images, this gap confirms the benchmark tests whether a method sees what the user shows.

{\header{Targets are also predictable from the user's history} SASRec and VBPR, which ignore the query and rank only from the history, reach an NDCG@1 of 0.41 to 0.44 on movies, far above the 0.23 of Popularity and close to the 0.43 to 0.46 of Direct MLLM. Because each target in \bench{} is an item the user later rated highly, the benchmark also tests whether a method sees the user.\par}

\begin{figure}[t]
\centering
\captionsetup{skip=4pt}
\includegraphics[width=\columnwidth,trim=0 5 0 3,clip]{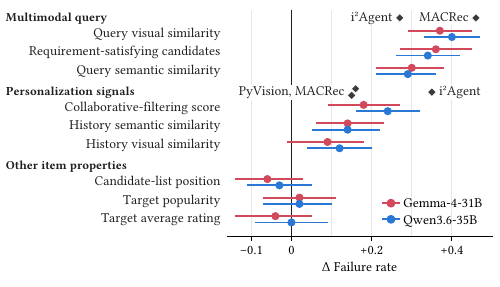}\par
\vspace*{-1mm}
\caption{Mean associations between nine factors and failures to rank the target first across the six agents on the random pool. Diamonds mark significant deviations from the mean (95\% bootstrap interval).}
\Description{Failures are associated with factors from the query requirements far more than with factors from the personalization signals.}
\label{fig:factors}
\vspace*{-3.5mm}
\end{figure}

\header{Agents reason more from query factors than from personalization signals} From the inputs of each test case, we extract nine factors that the agents may rely on, drawn from the multimodal query, the personalization signals, or other item properties, and measure how strongly every one is associated with failures to rank the target first~\cite{lyu2023featuredebiasing}. As shown in Figure~\ref{fig:factors} (factors defined in Appendix~\ref{app:factors}), the failure rate rises by up to 0.40 as the target resembles the reference images less, but by at most 0.24 for the strongest personalization factor, the collaborative-filtering score of the target. With both information sources available, these associations suggest that agents rely more on query-related factors than on personalization signals on the random pool.
\par\endgroup

\providecommand{\dn}[1]{{\color{red!70!black}\tiny$\downarrow$#1}}
\begin{table}[t]
\caption{Each agent's performance on the diagnostic pool with Gemma-4-31B. Red numbers are NDCG drops from the random pool. Best results are in bold; second best are underlined.}
\label{tab:pools}
\vspace*{-2mm}
\centering
\setlength{\tabcolsep}{2pt}\renewcommand{\arraystretch}{0.8}
\resizebox{\columnwidth}{!}{%
\begin{tabular}{l ccccc ccccc}
\toprule
& \multicolumn{5}{c}{Clothing} & \multicolumn{5}{c}{Movies} \\
\cmidrule(lr){2-6}\cmidrule(lr){7-11}
Method & @1 & @3 & @5 & $U$ & $P$ & @1 & @3 & @5 & $U$ & $P$ \\
\midrule
RecMind & \makecell{\underline{0.327}\\[-4.5pt]\dn{0.216}} & \makecell{\underline{0.494}\\[-4.5pt]\dn{0.171}} & \makecell{\underline{0.574}\\[-4.5pt]\dn{0.127}} & \underline{0.94} & 0.74 & \makecell{\underline{0.440}\\[-4.5pt]\dn{0.227}} & \makecell{\textbf{0.619}\\[-4.5pt]\dn{0.156}} & \makecell{\textbf{0.675}\\[-4.5pt]\dn{0.118}} & \underline{0.87} & \textbf{0.84} \\
InteRecAgent & \makecell{\textbf{0.332}\\[-4.5pt]\dn{0.218}} & \makecell{\textbf{0.516}\\[-4.5pt]\dn{0.137}} & \makecell{\textbf{0.579}\\[-4.5pt]\dn{0.126}} & \underline{0.94} & \underline{0.75} & \makecell{\textbf{0.451}\\[-4.5pt]\dn{0.196}} & \makecell{0.610\\[-4.5pt]\dn{0.170}} & \makecell{\underline{0.663}\\[-4.5pt]\dn{0.137}} & 0.85 & 0.81 \\
ReAct & \makecell{\textbf{0.332}\\[-4.5pt]\dn{0.246}} & \makecell{0.488\\[-4.5pt]\dn{0.193}} & \makecell{0.562\\[-4.5pt]\dn{0.150}} & 0.93 & \textbf{0.76} & \makecell{0.412\\[-4.5pt]\dn{0.247}} & \makecell{0.602\\[-4.5pt]\dn{0.179}} & \makecell{0.655\\[-4.5pt]\dn{0.144}} & 0.85 & \underline{0.83} \\
PyVision & \makecell{0.271\\[-4.5pt]\dn{0.237}} & \makecell{0.440\\[-4.5pt]\dn{0.193}} & \makecell{0.526\\[-4.5pt]\dn{0.146}} & \textbf{0.95} & 0.71 & \makecell{0.423\\[-4.5pt]\dn{0.260}} & \makecell{\underline{0.616}\\[-4.5pt]\dn{0.182}} & \makecell{0.658\\[-4.5pt]\dn{0.161}} & \textbf{0.88} & \underline{0.83} \\
\bottomrule
\end{tabular}}\par
\vspace*{-3.5mm}
\end{table}

\begin{figure}[!t]
\centering
\captionsetup{skip=4pt}
\includegraphics[width=0.93\columnwidth,trim=0 0 0 3,clip]{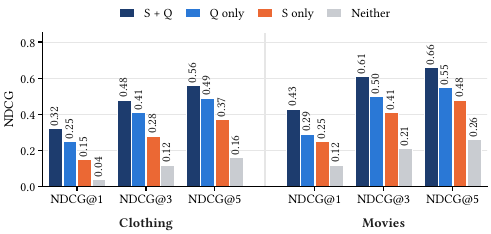}
\par\vspace{-3pt}{\footnotesize (a)}\par\vspace{-1pt}
\includegraphics[width=0.93\columnwidth,trim=0 0 0 3,clip]{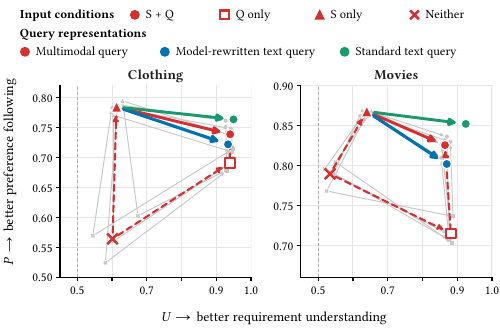}
\par\vspace{-3pt}{\footnotesize (b)}\par
\vspace*{-2mm}
\caption{Input ablations on the diagnostic pool with Gemma-4-31B, averaged over the four agents. (a) Ranking performance under four input conditions. (b) Requirement understanding ($U$) and preference following ($P$) under different inputs; gray paths show individual agents, and dashed arrows show the other input transitions. S and Q denote the personalization signals and the multimodal query.}
\Description{Panel (a): removing either the multimodal query or the personalization signals lowers mean NDCG in both domains. Panel (b): in both domains, adding a query to the personalization signals lowers mean preference-following scores while raising mean requirement-understanding scores; the decrease is largest for the model-rewritten text query and smallest for the standard text query.}
\label{fig:inputs}
\label{fig:four-inputs}
\label{fig:up}
\vspace*{-7mm}
\end{figure}

\vspace*{-3mm}
\subsection{Preference Following Analysis (RQ2)}
\label{sec:balance}
\begingroup

{Table~\ref{tab:pools} and Figure~\ref{fig:inputs} give the diagnostic-pool results with Gemma-4-31B. We analyze ReAct, RecMind, InteRecAgent, and PyVision, and exclude MACRec and \iiagent{}, whose multi-stage or fixed workflows make it hard to separate a model's own errors from errors between components. Agents understand the requirements well, but adding the multimodal query weakens their preference-following performance. We have three main observations:\par}

{\header{All agents drop sharply on the diagnostic pool} As shown in Table~\ref{tab:pools}, the diagnostic pool lowers the NDCG@1 of every agent by about 0.20 to 0.26 in both domains, and NDCG@3 and @5 also drop. Yet agents perform well on the random pool despite relying more on query-related factors than on personalization signals, suggesting that they rule out most random candidates largely through requirement understanding, whereas the competitors in \bench{} also require following the user's preferences. Thus, the random pool alone cannot fully evaluate personalization under multimodal queries, which establishes the need for our diagnostic pool.\par}

\header{Preference following leaves more room for improvement.} The four agents reach a $U$ of 0.85 to 0.95, ranking the target above most preference-aligning competitors, but their $P$ is only 0.71 to 0.84, so requirement-satisfying competitors still often outrank the target. Because \bench{} scores the two groups of competitors separately, it exposes that strong requirement understanding does not ensure personalized recommendations.

 {\header{The multimodal query helps ranking but crowds out personalization} We remove the personalization signals (S), the multimodal query (Q), or both. As shown in Figure~\ref{fig:inputs}(a), removing either one significantly lowers NDCG@1 in both domains. Figure~\ref{fig:inputs}(b) shows, however, that agents already achieve a high $P$ with S only, and adding Q significantly lowers the mean $P$ by 0.046 on clothing and 0.042 on movies (95\% CIs for $\Delta P$: $[-0.088, -0.005]$ and $[-0.078, -0.006]$; 2,000 user-level paired bootstrap resamples). This drop holds consistently for every agent, domain, and backbone. \textcolor{black}{With the personalization signals and the competitors held fixed in \bench{}, this drop can be attributed to the multimodal query, suggesting that it crowds out the personalization ability~\cite{lyu2025sacd} that agents already have and reveals a personalization imbalance.}\par}
\par\endgroup

\subsection{Multimodal Query Analysis (RQ3)}
\label{sec:representation}
\begingroup

\begin{figure*}[t]
\centering
\captionsetup{skip=4pt}
\begin{minipage}[t]{0.423\textwidth}
\centering
\includegraphics[width=\linewidth,trim=0 0 0 2.5,clip]{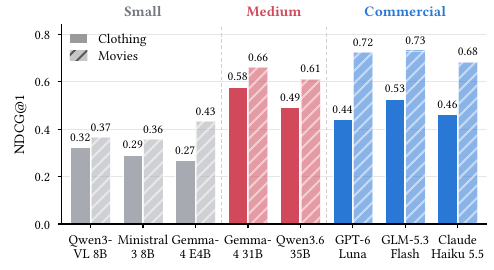}
\par\vspace{-3pt}{\footnotesize (a)}
\end{minipage}
\begin{minipage}[t]{0.459\textwidth}
\centering
\includegraphics[width=\linewidth,trim=0 0 0 3,clip]{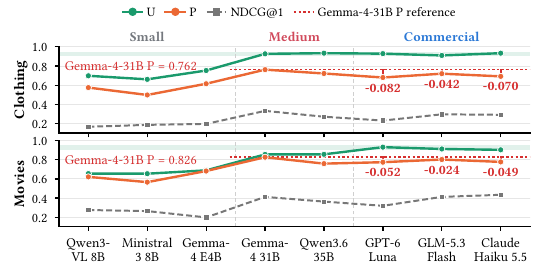}
\par\vspace{-3pt}{\footnotesize (b)}
\end{minipage}\par
\vspace*{-2mm}
\caption{Comparison of eight backbone models with the agent fixed to ReAct. (a) NDCG@1 on the random pool. (b) $U$, $P$, and NDCG@1 on the diagnostic pool; red numbers give the difference in $P$ from Gemma-4-31B.}
\Description{Panel (a): on the random pool, medium open-weight and commercial backbones score clearly higher than the small open-weight models. Panel (b): on the diagnostic pool, commercial backbones reach high requirement understanding but lower preference following than Gemma-4-31B in both domains.}
\label{fig:backbone-random}
\label{fig:model-ladder}
\vspace*{-3mm}
\end{figure*}

\begin{figure}[t]
\centering
\captionsetup{skip=4pt}
\begin{minipage}[t]{1.77in}
\centering
\includegraphics[width=\linewidth]{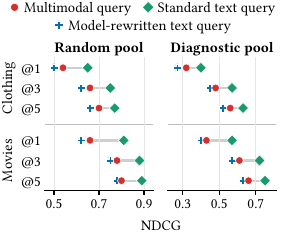}
\par\vspace{-3pt}{\footnotesize (a)}
\end{minipage}
\input{figures/f07_newly_first}\par
\vspace*{-2mm}
\caption{Comparison of query representations with Gemma-4-31B, averaged over the four agents. (a) Recommendation performance with full input. (b) NDCG@1 on the diagnostic pool with the query alone (circles) and after adding the personalization signals (bars); dark segments show targets promoted to first place.}
\Description{Panel (a) compares ranking performance under standard text, multimodal, and model-rewritten text queries across domains and pools. In panel (b), open circles mark query-only NDCG@1, while bar totals mark NDCG@1 after adding personalization signals. Dark segments show targets promoted from below first to first; light segments show targets ranked first both before and after. Targets that lose first place do not contribute to the bars.}
\label{fig:query-forms}
\label{fig:three-forms}
\label{fig:newly-first}
\vspace*{-3mm}
\end{figure}

Figure~\ref{fig:inputs}(b) and Figure~\ref{fig:query-forms} compare the multimodal query with two text-only representations on Gemma-4-31B, averaged over the four agents, with other inputs unchanged. The \emph{standard text query} states the attribute values of the requirements directly, and the \emph{model-rewritten text query} is written by the same backbone from the query text and reference images. Interpreting the reference images weakens agents' preference-following performance, yet keeping the images helps the history inform ranking. We have two observations:

{\header{Interpreting the reference images weakens preference-fol\-low\-ing performance} As shown in Figure~\ref{fig:inputs}(b), adding the standard text query to S only no longer significantly lowers $P$ in either domain and yields the highest NDCG in Figure~\ref{fig:query-forms}(a), whereas the model-rewritten text query lowers $P$ more than the multimodal query does, by 0.066 versus 0.042 on movies, with the same pattern on clothing. Both variants are text, but only the rewritten one depends on how the model reads the images, so this contrast separates image interpretation from image modality and shows that interpretation contributes to the loss of preference following.\par}

\header{Personalization signals help more when the reference images are kept} As shown in Figure~\ref{fig:query-forms}(a), with S, the multimodal query outperforms the model-rewritten text query in NDCG on both pools and domains. Without S, as the circles in Figure~\ref{fig:query-forms}(b) show, it still scores higher on clothing (25.1\% vs.\ 21.7\% in NDCG@1), suggesting that rewriting discards useful image information, but slightly lower on movies (28.7\% vs.\ 30.2\%), where the advantage of the multimodal query emerges only after the personalization signals are added. Figure~\ref{fig:query-forms}(b) shows that adding personalization signals promotes more targets to first place with the multimodal query, significantly on movies (21.0\% vs.\ 15.7\%) but not on clothing, consistent with LLMs' potentially richer pretrained knowledge~\cite{lyu2024knowtuning} of digital content than of physical products~\cite{dipalma2025memorize,tsai2024leveraging} and with agents rarely inspecting clothing history items. Because both representations receive the same S in \bench{}, this gain can be attributed to the reference images, which leave more room for the history to inform how agents read the query and search candidates.
\par\endgroup

\begin{figure*}[t]
\centering
\captionsetup{skip=4pt}
\includegraphics[width=0.873\textwidth,trim=0 0 0 1.5,clip]{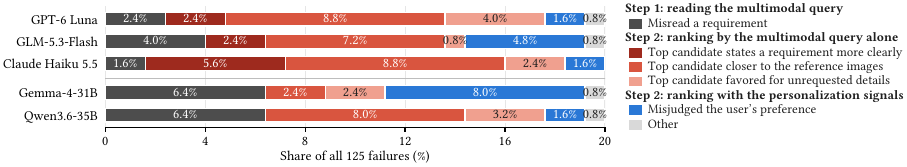}\par
\vspace*{-1mm}
\caption{First decisive errors in 125 sampled failures (25 per model); all percentages use 125 as the denominator.}
\Description{After reading the query correctly, most failures rank by query information alone.}
\label{fig:failure-types}
\vspace*{-3mm}
\end{figure*}

\vspace*{-4mm}
\subsection{Backbone Comparison (RQ4)}
\label{sec:backbone}
\begingroup

Figure~\ref{fig:backbone-random} compares the eight backbone models on both pools, with the agent fixed to ReAct. Stronger backbones improve requirement understanding, but the commercial models follow preferences less well than Gemma-4-31B. We have two main observations:

\header{Stronger backbones rank random candidates better through requirement understanding} In Figure~\ref{fig:backbone-random}(a), the medium open-weight and commercial models outperform the small open-weight models in NDCG@1, with 0.44 to 0.58 versus 0.27 to 0.32 on clothing and 0.61 to 0.73 versus 0.36 to 0.43 on movies. Since agents can rule out most random candidates in \bench{} by the query alone, these scores mainly follow $U$ in Figure~\ref{fig:model-ladder}(b): on movies, the commercial models exceed Gemma-4-31B in $U$ and outperform both medium models, whereas on clothing they only match Gemma-4-31B in $U$ and fall below it in NDCG@1 (0.44 to 0.53 versus 0.58), consistent with their weaker preference-following performance.

\header{Commercial models reach near-ceiling requirement understanding but weaker preference-following performance} The commercial models reach a $U$ of 0.90 to 0.93, close to saturation, yet their $P$ of 0.68 to 0.72 on clothing and 0.77 to 0.80 on movies stays below the 0.76 and 0.83 of Gemma-4-31B. On the diagnostic pool, the weaker $P$ offsets the gains in $U$, so the commercial models show no consistent NDCG@1 advantage over Gemma-4-31B on movies and score lower on clothing. Because \bench{} scores requirement understanding and preference following separately, this offset is visible, showing that, with ReAct, upgrading the backbone alone does not resolve the imbalance between the two.
\par\endgroup

\vspace*{-1mm}
\subsection{Failure Analysis (RQ5)}
\label{sec:failure}
\begingroup

Figure~\ref{fig:failure-types} presents the first decisive error in 25 randomly sampled failures to rank the target first for each commercial and medium open-weight model on the diagnostic pool, located in the execution traces by Claude Opus~5.5~\cite{cemri2025mast,merrill2026terminalbench,zhang2025whowhen}. Since agents first read the query and then rank, a failure is either a misread, where the agent omits an annotated requirement, or a ranking failure, where a misleading signal lets the top candidate beat the target. Cohen's kappa is 0.87 between these labels and an independent relabeling by GPT-5.6 Sol, and 0.80 between the model labels and human labels. Most failures occur not in reading the query but in ranking, where agents rely on the query alone and shift weight away from user preferences. We have two main observations:

\header{Most failures come from ranking by the query alone} As shown in Figure~\ref{fig:failure-types}, misreads account for 20.8\% of all 125 sampled failures. After correctly identifying the query requirements, 58.4\% of all failures come from ranking by query information alone: the top candidate looks closer to the reference images in 35.2\%, in 12.8\%, the agent uses unrequested details to further distinguish candidates that appear to match the query, and in 10.4\%, both candidates satisfy the requirements but the agent picks the one whose title or recalled facts state a requirement more clearly, an error made only by the commercial models. Only 17.6\% occur even though the agent combines the query with the personalization signals, and none comes from ranking by the personalization signals alone. Since the annotated requirements in \bench{} separate misreadings from ranking errors, this breakdown shows that agents can correctly identify the query requirements yet rely on query-related cues alone in subsequent ranking, without incorporating personalization signals.

{\header{Agents shift weight away from user preferences during ranking} In 59\% of these query-only ranking failures, the top candidate has a lower collaborative-filtering score than the target, contradicting an available personalization signal. In the execution traces, we find that agents sometimes even notice that their reasoning about ranking conflicts with user preferences yet keep the ranking unchanged. We show one such case in Appendix~\ref{app:case-clothing}. Although \bench{} provides user histories and preference evidence consistent with the query requirements, the agents still fail to incorporate this information into their final candidate rankings.\par}
\par\endgroup

\vspace*{-1mm}
\section{Related Work}
\label{sec:related}

{Existing benchmarks for recommendation with user requests fall into two categories. \textbf{(i) Benchmarks with textual queries.}~\cite{lyu2025deepshop} JDsearch and KuaiSearch pair real search queries with interaction feedback~\cite{liu2023jdsearch,li2026kuaisearch}, while PersonalWAB, InstructRec, and RecBench+ generate requests from user records~\cite{cai2025personalwab,xu2025iagent,huang2026recbench}. APeB combines underspecified queries with real user histories, separately evaluates intermediate intent and preference reasoning, and assesses final recommendations against actual purchases~\cite{yang2026apeb}. Persona2Web tests implicit preference inference in web tasks using synthetic browsing histories~\cite{kim2026persona2web}. Real histories tie evaluation to observed behavior, whereas generated histories test consistency with preferences encoded in synthetic personas. Despite differences in history provenance and evaluation targets, these textual-query settings do not test how interpreting reference images affects historical preference following.\par}

\textbf{(ii) Benchmarks with multimodal queries.} Composed image retrieval combines reference images with modification text to specify target relevance, without individual user histories~\cite{wu2021fashioniq,liu2021cirr,baldrati2023circo}. Imagery as Inquiry uses community-endorsed recommendations as targets~\cite{yoon2024imagery}, while MMShopBench verifies products against expert-annotated requirements~\cite{hao2026mmshopbench}. These settings assess request satisfaction without grounding evaluation in the requester's past item interactions. VOGUE evaluates preference inference from dialogue against participants' post-conversation ratings~\cite{guo2026vogue}, without testing whether models can infer preferences from prior rating histories. FashionRec supplies history-derived summaries in queries~\cite{pang2025fashionm3}, giving models explicit preference descriptions as in other multimodal personalization settings~\cite{kim2025mmpb}. Following these descriptions does not establish whether models can recover the same preferences from the original behavioral records. \bench{} instead leaves preferences implicit in real histories, so success requires relating the multimodal request to evidence of the same user's earlier behavior.

{Relating multimodal requests to user-specific evidence has recently drawn interest beyond benchmarks. IntentLens, for example, grounds such requests with visual, user-memory, and item-attribute tools~\cite{chen2026intentlens}, yet a more accurate final ranking alone does not establish better historical preference following. Overall ranking can improve via better query satisfaction while preference following declines. \bench{} therefore evaluates requirement understanding and preference following separately using final rankings. The preference-following score tests whether agents favor preference-aligned items among those satisfying the query, so that better requirement understanding is not mistaken for better personalization.\par}

\vspace*{-1mm}
\section{Conclusion}
\label{sec:conclusion}
\label{sec:discussion}

\begingroup

In this paper, we focus on evaluating the personalization of recommendation agents under multimodal queries that combine text with reference images. We have presented \bench{}, a benchmark of 510 clothing and movie queries from real user records. Its diagnostic pools support separate measures of requirement understanding and preference following. Experiments show that both multimodal queries and personalization signals improve agents’ overall ranking performance. Yet adding the queries to personalization signals weakens preference following despite providing additional information about query requirements, revealing a personalization imbalance. Moreover, query-representation contrasts link this imbalance to the interpretation of the reference images, failure traces show that most failures involve multimodal query-based cues overriding preference evidence, and stronger backbones do not resolve it. These findings call for personalization-aware ranking~\cite{lyu2025macpo}, so that user preferences shape the final choice. The data and evaluation code are available at \url{https://github.com/zxd-octopus/PMRecBench-Public}.
\par\endgroup

\bibliographystyle{ACM-Reference-Format}
\bibliography{refs}

\appendix

\section{Query Construction Details}
\label{app:data}\label{app:data-sources}

\header{Requirement Mining} We select users with at least five liked items (ratings $\geq 4$) and five disliked items (ratings $\leq 2$).  Requirements need at least two distinct supporting items; a supporting clothing review must mention the attribute. Negative attributes must be over 0.10 more prevalent among disliked than liked items and occur in at most 5\% of liked items to identify dislike-associated attributes.

\header{Reference Image Retrieval} We retrieve movie stills with the Serper image API\footnote{\url{https://serper.dev}} using the title, release year, and attribute, excluding posters and covers to reduce textual clues about the attribute. Per requirement, we check images from up to three supporting items; if none passes, we treat the attribute as not visible. In Combined queries, one image may show both requirements.

\header{Query Generation} Drafting and paraphrasing use temperatures of 0.1 and 0.6, respectively. Queries contain one or two location anchors and at most two sentences: 10--25 words for Single and Shared queries, and 15--35 for Combined and Contrastive queries.

\vspace*{-1mm}
\section{Implementation Details}
\label{app:prompts}\label{app:methods}\label{app:tools}
\header{Task instructions} Each agent keeps its original prompt, to which we add a unified task instruction asking it to rank candidates by whether they satisfy the query requirements and match the user's preferences. The prompt below is for agents. For direct rankers, we remove the tool-use instructions and put candidate images, history titles, and history image captions directly in the prompt. Agent frameworks may coordinate specialized roles~\cite{wang2025cooperative}, accumulate reusable experience through persistent memory~\cite{lyu2026evoscientist}, or evolve through feedback shared across agents~\cite{meng2026swarmrsi}; these design choices are outside the controlled prompt change evaluated here.

\begin{figure}[h]
\leftskip=0pt\rightskip=0pt\parfillskip=0pt plus 1fil\relax
\noindent\begin{minipage}{\columnwidth}
\begingroup
\setlength{\parindent}{0pt}
\setlength{\FrameSep}{4pt}\setlength{\FrameRule}{0.4pt}
\begin{framed}\footnotesize
\noindent\textbf{Task Instruction for Agents}\par\nopagebreak\smallskip
Task: a user is looking for a \texttt{\{movie/clothing\}} recommendation.\par\smallskip
You are given:
\setlength{\leftmargini}{1.2em}\setlength{\topsep}{2pt}\setlength{\partopsep}{0pt}%
\begin{itemize}\setlength{\itemsep}{0pt}\setlength{\parskip}{0pt}
\item the user's request consists of text plus one or more reference images, numbered ``Query image 1'', ``Query image 2'', \ldots{}; these images are part of the request;
\item a fixed list of candidates is provided, each shown only as [id] title;
\item the user's preference signals are NOT given here; you must fetch them yourself by calling tools;
\item all items (including history items and candidates) have ids; use these ids to retrieve their images, captions, metadata, etc., through tools.
\end{itemize}
Output a ranking of ALL candidate ids (use the bracketed ids exactly), from most to least suitable. The most suitable \texttt{\{movie/clothing\}} must both satisfy the request (its text and its reference images) and align with the user's preference signals.\par\smallskip
\textbf{Input:}\par
User request: \texttt{\{query\_text\}}\\
Reference images: \texttt{\{numbered\_images\}}\\
Candidate list:
\texttt{\{candidate\_list\}}\par
\end{framed}
\endgroup
\end{minipage}
\vspace*{-3.5mm}
\end{figure}

\header{Item information} During evaluation, we mask attributes used to construct each query in all candidate profiles and profiles of history items supplying its reference images, to prevent information leakage. Each candidate image is a collage of up to three images: the hero image followed by other photos of the same item for clothing, or the main poster followed by stills for movies. Each backbone generates its own image captions. Each query has a fixed candidate order shared by all methods and runs.

\header{Score computation} The collaborative-filtering score averages the pool-standardized scores of SASRec and VBPR; query users contribute only the earlier 70\% of their records to training. CLIP similarities use the openai/clip-vit-base-patch32 model.

\providecommand{\dn}[1]{{\color{red!70!black}\tiny$\downarrow$#1}}
\begin{table}[t]
\caption{Performance of each agent on the diagnostic pool with Qwen3.6-35B, in the same format as Table~\ref{tab:pools}.}
\label{tab:pools-qwen}
\centering
\setlength{\tabcolsep}{2pt}\renewcommand{\arraystretch}{0.8}
\resizebox{\columnwidth}{!}{%
\begin{tabular}{l ccccc ccccc}
\toprule
& \multicolumn{5}{c}{Clothing} & \multicolumn{5}{c}{Movies} \\
\cmidrule(lr){2-6}\cmidrule(lr){7-11}
Method & @1 & @3 & @5 & $U$ & $P$ & @1 & @3 & @5 & $U$ & $P$ \\
\midrule
RecMind & \makecell{\textbf{0.317}\\[-4.5pt]\dn{0.202}} & \makecell{\textbf{0.469}\\[-4.5pt]\dn{0.172}} & \makecell{\underline{0.536}\\[-4.5pt]\dn{0.124}} & \textbf{0.94} & \textbf{0.74} & \makecell{\underline{0.379}\\[-4.5pt]\dn{0.216}} & \makecell{0.525\\[-4.5pt]\dn{0.206}} & \makecell{\underline{0.598}\\[-4.5pt]\dn{0.154}} & \underline{0.87} & 0.77 \\
InteRecAgent & \makecell{\underline{0.301}\\[-4.5pt]\dn{0.211}} & \makecell{0.446\\[-4.5pt]\dn{0.190}} & \makecell{0.509\\[-4.5pt]\dn{0.163}} & 0.91 & 0.68 & \makecell{\underline{0.379}\\[-4.5pt]\dn{0.296}} & \makecell{\textbf{0.578}\\[-4.5pt]\dn{0.202}} & \makecell{\textbf{0.640}\\[-4.5pt]\dn{0.161}} & \textbf{0.88} & \textbf{0.80} \\
ReAct & \makecell{0.271\\[-4.5pt]\dn{0.221}} & \makecell{0.437\\[-4.5pt]\dn{0.176}} & \makecell{0.517\\[-4.5pt]\dn{0.143}} & \underline{0.93} & \underline{0.72} & \makecell{0.363\\[-4.5pt]\dn{0.248}} & \makecell{0.526\\[-4.5pt]\dn{0.185}} & \makecell{0.587\\[-4.5pt]\dn{0.158}} & 0.86 & 0.76 \\
PyVision & \makecell{\underline{0.301}\\[-4.5pt]\dn{0.187}} & \makecell{\underline{0.461}\\[-4.5pt]\dn{0.138}} & \makecell{\textbf{0.545}\\[-4.5pt]\dn{0.095}} & 0.92 & \underline{0.72} & \makecell{\textbf{0.412}\\[-4.5pt]\dn{0.239}} & \makecell{\underline{0.577}\\[-4.5pt]\dn{0.184}} & \makecell{\textbf{0.640}\\[-4.5pt]\dn{0.135}} & \textbf{0.88} & \underline{0.79} \\
\bottomrule
\end{tabular}}
\vspace*{-3.5mm}
\end{table}

\begin{figure}[t]
\centering
\captionsetup{skip=4pt}
\includegraphics[width=0.82\columnwidth,trim=0 0 0 3,clip]{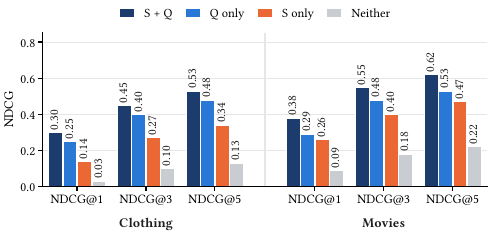}
\par\vspace{-3pt}{\footnotesize (a)}\par\vspace{-1pt}
\includegraphics[width=0.82\columnwidth,trim=0 0 0 3,clip]{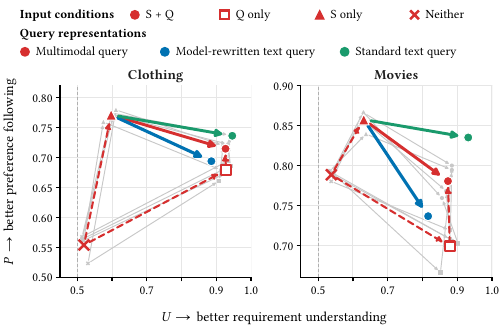}
\par\vspace{-3pt}{\footnotesize (b)}\par
\input{figures/f05_three_forms_qwen}\par
\vspace*{-1mm}
\caption{Qwen3.6-35B results. Panels (a--b) and (c--d) follow the formats of Figures~\ref{fig:inputs} and~\ref{fig:query-forms}, respectively.}
\Description{Panel (a): removing either the multimodal query or the personalization signals lowers mean NDCG in both domains. Panel (b): adding a query to the personalization signals lowers mean preference-following scores while raising mean requirement-understanding scores. Panel (c) compares ranking performance under standard text, multimodal, and model-rewritten text queries across domains and pools. In panel (d), open circles mark query-only NDCG@1, while bar totals mark NDCG@1 after adding personalization signals. Dark segments show targets promoted from below first to first; light segments show targets ranked first both before and after. Targets that lose first place do not contribute to the bars.}
\label{fig:qwen-results}
\label{fig:inputs-qwen}
\label{fig:four-inputs-qwen}
\label{fig:up-qwen}
\label{fig:query-forms-qwen}
\label{fig:three-forms-qwen}
\label{fig:rank-change-qwen}
\vspace*{-3.5mm}
\end{figure}

\section{Results with Qwen3.6-35B}
\label{app:results}\label{app:qwen-balance}
The main trends with Qwen3.6-35B are consistent with those in Sections~\ref{sec:balance} and~\ref{sec:representation} (Table~\ref{tab:pools-qwen} and Figure~\ref{fig:qwen-results}).

\section{Input Factor Analysis}
\label{app:factors}
We define the factors and calculation behind the failure-rate comparisons in Figure~\ref{fig:factors} (Section~\ref{sec:overall}).
{A failure is a run in which the target is not ranked first. Three factors come from the query: \emph{query visual similarity} between the target image and the reference images, \emph{query semantic similarity} between their image captions, and the number of other \emph{requirement-satisfying candidates}. Three come from the personalization signals: the \emph{collaborative-filtering score} of the target, its maximum similarity to the user's history items based on titles and metadata (\emph{history semantic similarity}), and its maximum image similarity to those items (\emph{history visual similarity}). Three are other properties: \emph{candidate-list position}, \emph{target popularity}, and \emph{target average rating}. Visual similarity is the cosine similarity of CLIP image embeddings~\cite{radford2021clip}, and semantic similarity that of sentence embeddings~\cite{reimers2019sbert}.
For each factor, we split the test cases of each domain into thirds by the target's value (for requirement-satisfying candidates, zero versus three or more) and report the failure rate of the worst third minus that of the best third, pooling the six agents and both domains. Diamonds in Figure~\ref{fig:factors} mark agents whose difference, pooled over the two backbones, differs significantly from the six-agent mean.\par}

\begin{table}[t]
\caption{NDCG@1/@3/@5 of ReAct on the random pool, averaged within each model group. Contrastive is shaded.}
\label{tab:qtype}
\centering
\setlength{\tabcolsep}{3pt}\renewcommand{\arraystretch}{0.9}
\resizebox{\columnwidth}{!}{%
\begin{tabular}{l cccc}
\toprule
& Single & Shared & Combined & Contrastive \\
\midrule
\multicolumn{5}{l}{\textit{Small open-weight (Qwen3-VL-8B, Ministral 3 8B, Gemma-4-E4B)}} \\
Clothing & 0.32/0.39/0.43 & 0.31/0.39/0.42 & 0.32/0.41/0.46 & \cellcolor{hlRed}0.23/0.32/0.36 \\
Movies & 0.31/0.43/0.46 & 0.45/0.56/0.58 & 0.45/0.58/0.61 & \cellcolor{hlRed}0.33/0.43/0.47 \\
\midrule
\multicolumn{5}{l}{\textit{Medium open-weight (Gemma-4-31B, Qwen3.6-35B)}} \\
Clothing & 0.58/0.66/0.69 & 0.56/0.65/0.70 & 0.56/0.70/0.73 & \cellcolor{hlRed}0.44/0.58/0.63 \\
Movies & 0.65/0.74/0.77 & 0.66/0.78/0.81 & 0.74/0.81/0.84 & \cellcolor{hlRed}0.50/0.65/0.68 \\
\midrule
\multicolumn{5}{l}{\textit{Commercial (GPT-6 Luna, GLM-5.3-Flash, Claude Haiku 5.5)}} \\
Clothing & 0.53/0.65/0.68 & 0.46/0.58/0.62 & 0.56/0.70/0.72 & \cellcolor{hlRed}0.36/0.52/0.58 \\
Movies & 0.73/0.82/0.84 & 0.80/0.87/0.89 & 0.77/0.87/0.88 & \cellcolor{hlRed}0.55/0.70/0.72 \\
\bottomrule
\end{tabular}}
\vspace*{-3.5mm}
\end{table}

\begin{figure*}[t]
\centering
\includegraphics[width=\textwidth,trim=7.5 40 8 6,clip]{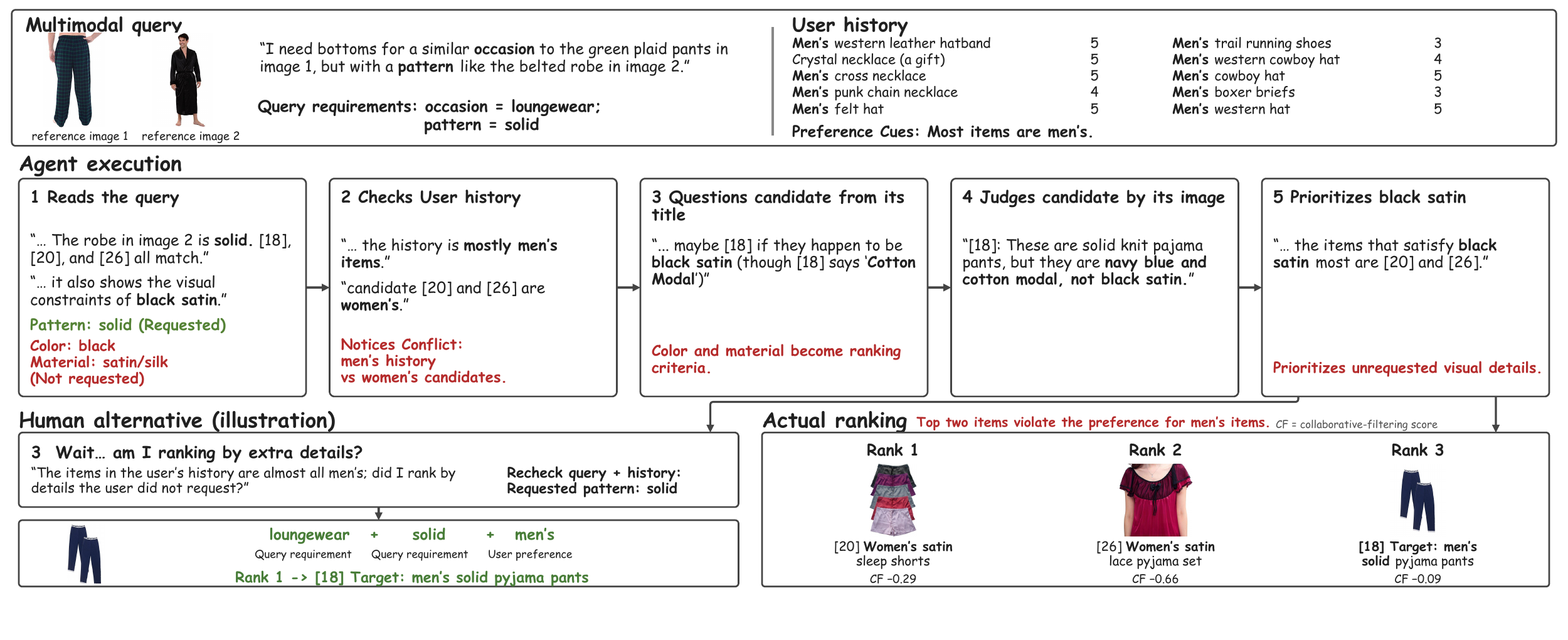}
\caption{Clothing case (ReAct, GLM-5.3-Flash, diagnostic pool). \emph{Agent execution}: selected steps and excerpts from the actual execution trace. \emph{Actual ranking}: the run's top three. \emph{Human alternative}: an illustrative correction.}
\Description{The text requests an occasion from the pants image and a pattern from the robe image. The agent also adds black color and satin or silk material. During ranking, it notices that the history mostly contains men's items and that candidates 20 and 26 are women's items, but retains the extra constraints and ranks them first and second ahead of the men's target 18. An illustrative human path uses this conflict to reconsider the added constraints and select target 18.}
\label{fig:case-clothing}
\vspace*{-2mm}
\end{figure*}

\vspace*{-1.5mm}
\section{Ranking across Query Forms}
\label{app:supp}
\label{app:forms-results}
{Table~\ref{tab:qtype} compares ranking performance across the four query forms of Section~\ref{sec:query-generation} on the random pool, using ReAct with multimodal queries and personalization signals on eight backbones: Qwen3-VL-8B, Ministral 3 8B, Gemma-4-E4B, Gemma-4-31B, Qwen3.6-35B, GPT-6 Luna, GLM-5.3-Flash, and Claude Haiku 5.5. Contrastive generally scores lowest, whereas Combined generally beats Single, indicating that extra query conditions help agents rule out random candidates that meet only part of the request in this ranking task.\par}

\vspace*{-1mm}
\section{Case Study}
\label{app:case-clothing}
Figure~\ref{fig:case-clothing} illustrates one of the failure types discussed in Section~\ref{sec:failure}: the top candidate is favored for unrequested details even after the agent recognizes a preference conflict.

{\header{The agent adds attributes that the query does not request} The text specifies the attribute type to read from each image: the pants' occasion (loungewear) and the robe's pattern (solid). The agent also ranks by black color and satin/silk material. This changes the ranking criterion from satisfying the requested attributes to resembling the reference image, penalizing candidates for unrequested visual details without giving user preference sufficient weight~\cite{lyu2024cognitivebiases}.\par}

{\header{The preference conflict does not trigger correction in ranking} The agent ranks candidates ``[20]'' and ``[26]'', both women's items, first and second for their black satin appearance, placing the men's target ``[18]'' third. The history predominantly contains men's items, providing a preference cue for men's clothing. Although the agent notices this conflict, it keeps its ranking unchanged, still favoring unrequested appearance details over the preference evidence. A possible human correction is to discard the unrequested ranking criteria when the preference conflict emerges. \par}

\vspace*{-1mm}
\section{Tool-Call Efficiency Analysis}
\label{app:cost}
\begin{table}[t]
\caption{Mean tool calls per test case on the diagnostic pool with S only and with S + Q.}
\label{tab:cost-sq}
\centering\footnotesize\setlength{\tabcolsep}{3pt}\renewcommand{\arraystretch}{0.85}
\begin{tabular*}{\columnwidth}{@{\extracolsep{\fill}}llcccc@{}}
\toprule
& & \multicolumn{2}{c}{Clothing} & \multicolumn{2}{c}{Movies} \\
\cmidrule(lr){3-4}\cmidrule(l){5-6}
Backbone & Agent & S only & S + Q & S only & S + Q \\
\midrule
\multirow{4}{*}{Gemma-4-31B} & ReAct & 2.26 & 3.75 & 2.95 & 4.15 \\
 & RecMind & 2.87 & 4.16 & 3.09 & 4.53 \\
 & InteRecAgent & 3.51 & 4.03 & 2.25 & 3.64 \\
 & PyVision & 4.22 & 4.16 & 4.05 & 5.80 \\
\midrule
\multirow{4}{*}{Qwen3.6-35B} & ReAct & 3.93 & 3.92 & 4.25 & 4.67 \\
 & RecMind & 5.68 & 5.89 & 5.35 & 7.02 \\
 & InteRecAgent & 2.54 & 3.38 & 2.57 & 3.58 \\
 & PyVision & 8.92 & 11.16 & 9.94 & 11.22 \\
\bottomrule
\end{tabular*}
\vspace*{-2mm}
\end{table}

\begin{figure}[t]
\centering
\includegraphics[width=1.0\columnwidth]{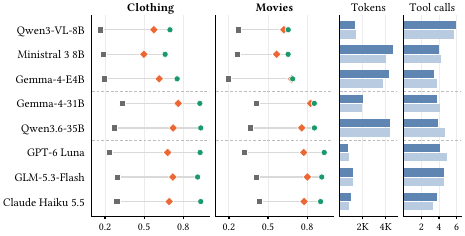}
\caption{NDCG@1 (squares), $P$ (diamonds), and $U$ (circles) of ReAct on the diagnostic pool, with mean output tokens and tool calls per test case (dark: clothing; light: movies).}
\Description{For each of the eight backbones, grouped into small open-weight, medium open-weight, and commercial models, the figure shows NDCG@1, P, and U on clothing and movies, and bars of mean output tokens and tool calls per test case. The models with the most output tokens or tool calls do not have higher NDCG@1, U, or P.}
\label{fig:cost-dots}
\vspace*{-3.5mm}
\end{figure}

Table~\ref{tab:cost-sq} reports the mean tool calls per test case of the four agents on the diagnostic pool with S only and with S + Q, and Figure~\ref{fig:cost-dots} compares the eight backbone models with ReAct. Unlike work on retrieving and reranking agents using execution-grounded signals~\cite{wu2026agentsearchbench} or accelerating calibrated within-agent search decisions~\cite{cao2026searchjev}, our analysis holds the agent fixed and measures how query inputs change its tool use.

\header{The decline in preference following is not caused by insufficient tool use} Adding the multimodal query to the personalization signals increases the mean tool calls per test case in 14 of the 16 settings of four agents, two backbones, and two domains, and leaves them nearly unchanged in the other two, yet lowers $P$ in all 16 settings. Agents therefore gather no less information under multimodal queries; the decline lies in how they use it when ranking, which link it to the interpretation of the reference images.

{%
\section{Validation of Filled Requirement-Satisfying Competitors}
\label{app:filled}
\begin{table}[t]
\caption{$P$ on disliked and all requirement-satisfying competitors, pooled over the four agents. *: significant change (95\% bootstrap interval).}
\label{tab:filled}
\centering\footnotesize\setlength{\tabcolsep}{3pt}\renewcommand{\arraystretch}{0.85}
\begin{tabular*}{\columnwidth}{@{\extracolsep{\fill}}llcccccc@{}}
\toprule
& & \multicolumn{3}{c}{Disliked} & \multicolumn{3}{c}{All} \\
\cmidrule(lr){3-5}\cmidrule(l){6-8}
Backbone & Domain & S only & S + Q & Change & S only & S + Q & Change \\
\midrule
\multirow{2}{*}{Gemma-4-31B} & Clothing & 0.650 & 0.517 & $-$0.133* & 0.785 & 0.739 & $-$0.046* \\
 & Movies & 0.764 & 0.670 & $-$0.094* & 0.868 & 0.826 & $-$0.042* \\
\multirow{2}{*}{Qwen3.6-35B} & Clothing & 0.617 & 0.517 & $-$0.100* & 0.771 & 0.715 & $-$0.056* \\
 & Movies & 0.722 & 0.639 & $-$0.083* & 0.858 & 0.780 & $-$0.078* \\
\bottomrule
\end{tabular*}
\vspace*{-2mm}
\end{table}
{When fewer than five disliked items satisfy the query, Section~\ref{sec:candidates} fills $\DP$ with never-interacted items under an attribute-overlap cap to reduce the chance of matching the user's preferences. Across both domains, the disliked-only subset covers 14.70\% of queries and 15.90\% of users in the diagnostic pool. Table~\ref{tab:filled} compares $P$ against disliked competitors alone and against all competitors, pooled over ReAct, PyVision, RecMind, and InteRecAgent on the diagnostic pool. Adding Q significantly lowers $P$ in both evaluations across domains and backbones, with larger drops against disliked competitors; including filled competitors therefore gives a more conservative estimate of the loss. The larger decline against disliked competitors may reflect visual similarities between previously interacted items and the history-derived reference images, encouraging agents to prioritize visual matching over preference evidence. The decline within the observed-dislike subset supports using $P$ to assess preference following in a diagnostic pool containing filled requirement-satisfying competitors, although the subset's limited coverage constrains generalization.\par}}

\begin{table}[t]
\caption{Individual scores, equal-weight fusion, and agents. NDCG averages @1, @3, and @5. Agent results are shown as Gemma-4 / Qwen3.6, each averaged over ReAct, RecMind, InteRecAgent, and PyVision under S+Q.}
\label{tab:score-fusion}
\centering\footnotesize\setlength{\tabcolsep}{2pt}\renewcommand{\arraystretch}{0.85}
\begin{tabular*}{\columnwidth}{@{\extracolsep{\fill}}llcccc@{}}
\toprule
Domain & Method & \makecell{NDCG\\Random} & \makecell{NDCG\\Diagnostic} & $U$ & $P$ \\
\midrule
\multirow{4}{*}{Clothing} & CF-only & 0.30 & 0.18 & 0.54 & 0.70 \\
 & CLIP-only & 0.27 & 0.23 & 0.74 & 0.64 \\
 & Fusion & 0.36 & 0.24 & 0.68 & 0.72 \\
 & Agents & 0.63 / 0.59 & 0.45 / 0.43 & 0.94 / 0.93 & 0.74 / 0.72 \\
\midrule
\multirow{4}{*}{Movies} & CF-only & 0.62 & 0.40 & 0.64 & 0.88 \\
 & CLIP-only & 0.40 & 0.29 & 0.79 & 0.69 \\
 & Fusion & 0.69 & 0.48 & 0.75 & 0.88 \\
 & Agents & 0.75 / 0.72 & 0.57 / 0.52 & 0.86 / 0.87 & 0.83 / 0.78 \\
\bottomrule
\end{tabular*}
\vspace*{-2mm}
\end{table}
{%
\section{Score Fusion Baseline}
\label{app:score-fusion}
{We test whether simple score fusion can match recommendation agents (Table~\ref{tab:score-fusion}). Fusion uses $\alpha=0.5$ to average CLIP and CF scores after z-standardizing both within each candidate pool. CLIP uses maximum similarity to positive reference images without interpreting query requirements.
We found fusion achieves higher random-pool mean NDCG than either score alone. Relative to CF-only, $U$ is higher in both domains; $P$ is higher on clothing but slightly lower on movies, where both values round to 0.88. Agents reach higher mean NDCG in both pools and higher $U$ than fusion, but not consistently higher $P$. Their edge in overall ranking and requirement satisfaction therefore does not ensure stronger preference following.\par}

}

\clearpage
\end{document}